\documentclass[prl,aps,reprint]{revtex4-1}
\usepackage{graphicx}
\usepackage{amsmath,amssymb}
\usepackage[caption=false]{subfig}
\usepackage[breaklinks=true,colorlinks=true,bookmarks=false,urlcolor=blue,
citecolor=blue,linkcolor=blue]{hyperref}

\begin{document}

\title{Breakup of liquid jets 
and the formation of satellite and subsatellite droplets}
\author{Fei Wang$^{1}$}
\email{fei.wang@kit.edu}
\author{Oleg Tschukin$^{1}$}
\author{Gabriel Cadilha Marques$^{2,4}$}
\author{Michael Selzer$^{1,3}$}
\author{Jasmin Aghassi-Hagmann$^{4,5}$}
\author{Britta Nestler$^{1,3}$}
\email{britta.nestler@kit.edu}

\affiliation{$^{1}$Institute of Applied Materials-Computational Materials Science, 
Karlsruhe Institute of Technology (KIT),
Stra{\ss}e am Forum 7, 76131 Karlsruhe, Germany}

\affiliation{$^{2}$Department of Computer Science, Karlsruhe Institute of Technology (KIT),
Haid-und-Neu-Str. 7, 76131 Karlsruhe, Germany}

\affiliation{$^{3}$Institute of Digital Materials Science, 
Karlsruhe University of Applied Sciences,
Moltkestra{\ss}e 30, 76133 Karlsruhe, Germany}

\affiliation{$^{4}$Institute of Nanotechnology, Karlsruhe Institute of Technology (KIT),
Hermann-von-Helmholtz-Platz 1, 76344 Eggenstein-Leopoldshafen, Germany} 

\affiliation{$^{5}$Department of Electrical Engineering and Information Technology, 
Offenburg University of Applied Sciences,  77652 Offenburg, Germany}

  \date{\today}
\begin{abstract}
A spontaneous breakup of a liquid jet results 
in the formation of a chain of droplets, which is a daily observed phenomenon, 
such as in the raining process and under an open water-faucet. 
We here report inkjet printing experiments
for the formation of droplets which are inconsistent with existing
theories and thereafter propose a novel concept of surface area minimization
 to address the  droplet-formation condition.
The novel concept  is also capable of predicting 
the formation of satellite and subsatellite droplets,
which cannot be comprehended by the classic Rayleigh theory for the
formation of droplets. We also present a new evolution behavior
of liquid jets not yet reported before that the jet initially transforms into droplets 
and then becomes a uniform-radius jet again. The current results can be applied to a variety
of disciplines where liquid jets are present. 
\end{abstract}

\maketitle

When we turn on a water-tap to take water,
the water jet sometimes breaks up into 
a chain of droplets~\cite{petit2012break,fragkopoulos2015teaching,wheeler2012physics}. 
This breakup phenomenon is also observed
during raining.
The spontaneous breakup of a liquid jet into a droplet-structure has been broadly 
used in a variety of fields.
For instance, when an inkjet leaves the head-nozzle of a printer,  
the jet falls down and breaks into several droplets
prior to contact with printed targets~\cite{martin2008inkjet,singh2010,wang2014liquid,Wheeler2014}.
In blood vessels, the breakup forms a so-called ``sausage-string'' 
pattern~\cite{alstrom1999instability}.
The breakup of liquid jets has been also used in many 
other interdisciplinary topics~\cite{zhang2016hydrodynamic,suzuki2016wafer,eggers2008physics,cardoso2006rayleigh,passian2012,link2004geometrically,azuma2014hagedorn,baumchen2014influence,kong2002pulmonary,xia2007potential,liu2008dendrite,schwarz1994physical,mabille2003monodisperse,Duclaux2004}, such as fabrications of 
nanoparticles from nanowires~\cite{kaufman2012structured,karim2006morphological,ko2016hierarchical}, 
formations of Li-Pb droplets in nuclear fusion~\cite{okino2012vacuum}, 
and the coating process of polymer thin films~\cite{tsai2013rayleigh}.

\begin{figure}[b]
\centering
\includegraphics[width=8.5cm]{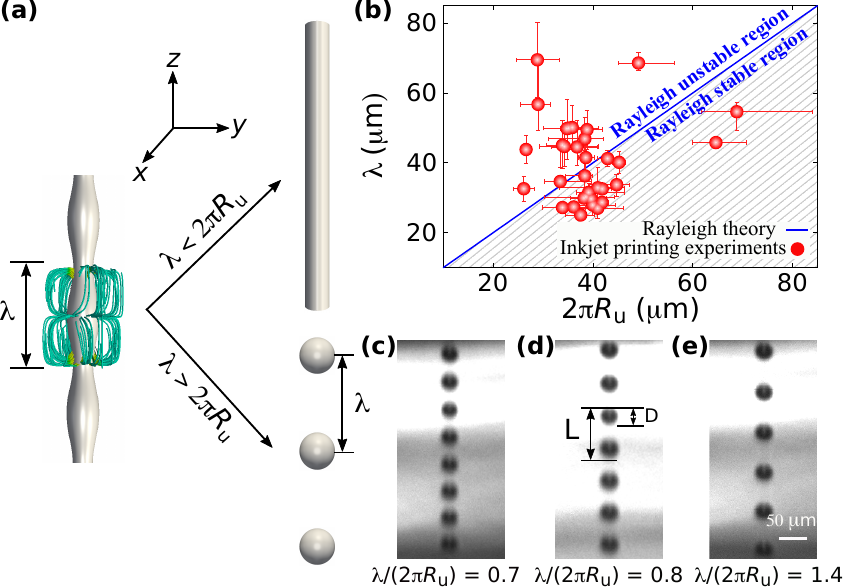}
\caption{Rayleigh's theory and the present inkjet printing experiments.
(a) Sketch of Rayleigh's theory.
The green lines are the stream lines of the capillary flow 
from our phase-field simulations. 
(b) Present inkjet printing experiments (red shaded circles) 
show that  droplets are observed in the Rayleigh's stable region (hatch lines) 
in which droplets should not form according to Rayleigh's theory. 
(c), (d) and (e) Droplet-structures from experiments with different
ratios of $\lambda/(2\pi R_u)$.  
Here, $\lambda$ is calculated by $L-D$ and $R_u$ is estimated 
by equivalencing the volume of the droplets  $\pi D^3/6$ to the
one of a uniform-radius jet $\pi R_u^2 \lambda$.}
	\label{fig:1}
\end{figure}

The classic droplet-formation criterion was proposed by 
Lord Rayleigh~\cite{Rayleigh1878,Rayleigh1879}, as sketched 
in  Fig.~\ref{fig:1}(a),
where a jet is axi-symmetrically
perturbed by a single cosine function, 
$r(z)=R_0 + a_1\cos (2 \pi z/\lambda) $, in the radial dimension.
Here, $R_0$ is a radius, $a_1$ is the amplitude and 
$\lambda$ is the wavelength.
According to Rayleigh's theory,
droplets are formed if and only if the wavelength $\lambda$ is greater than the 
circumference of the jet $2\pi R_u$.
However, we presently report inkjet printing
experiments~\footnote{See Supplemental Materials for the experimental description, 
the analysis and the phase-field model.} (Fig.~\ref{fig:1}(b))
that droplet-structures are formed in the 
Rayleigh morphological stable region
where droplets should not occur according to Rayleigh's theory.
Fig.~\ref{fig:1}(c)-(e) illustrate
three typical droplet-structures from the inkjet printing experiments,
where the droplet-formations in Fig.~\ref{fig:1}(c) and (d) 
are inconsistent with Rayleigh's theory.
In order to address this newly observed  phenomenon of droplet-formation,
a novel concept will be proposed in the present work.
Our new criterion not only generalizes the classic Rayleigh breakup
but also comprehends the formation of satellite droplets.
In addition, we find a new evolution behavior of liquid jets not yet reported in literature: 
the jet initially transforms into 
a droplet-structure and then becomes a uniform-radius cylinder once again.

 According to the physical principle 
of free energy minimization, 
the perturbed jet must evolve in such a way to reduce its
surface area irrespective of the evolution dynamics.
The surface area of the perturbed jet $A$ is calculated as:
$A=\int_0^\lambda 2\pi r\sqrt{1+(\partial_zr)^2} dz$
and the total surface energy is $\sigma A$~\footnote{Here, the actual value of surface energy density $\sigma$
is irrelevant for the calculation and is set to be unity for convenience.}.
In contrast to the work in literature~\cite{yuen1968non,nichols1965,nayfeh1970nonlinear}, 
two significant improvements are presently made  for the calculation of
the surface area landscape. (i) The calculation of 
the surface area must be subject to the condition 
that the volume of the jet is conserved, namely, 
$\int_0^\lambda \pi r^2 dz=\text{constant}=\pi R_u^2 \lambda$,
where $R_u$ is the radius of a uniform-radius jet
which has the equivalent volume as the perturbed jet.
This condition is essentially important 
and has not been considered in Rayleigh's original
work~\cite{Rayleigh1878}  as well as the corresponding linear stability analysis~\cite{nichols1965}.
(ii) The second significant fact is that 
although at the beginning only a single cosine perturbation $a_1\cos kz$ is considered,
the surface of the jet does not necessarily remain harmonic with time. 
The surface of the jet in general can be described by a Fourier series as 
\begin{equation}
r(z,a_1,a_2,\cdots)=R_0+a_1\cos kz+a_2\cos 2kz+\cdots,
\label{eq:1}
\end{equation}
where $a_i$, $i\in \mathbb N$  are the amplitudes of different orders
and $k$ is the wavenumber.
 By writing such a general expression for the jet, our consideration is not only restricted 
to the idealized case of Rayleigh where a single cosine deformation
is supposed to take place at the beginning.
\begin{figure}[b]
	\centering
	\includegraphics[width=1\linewidth]{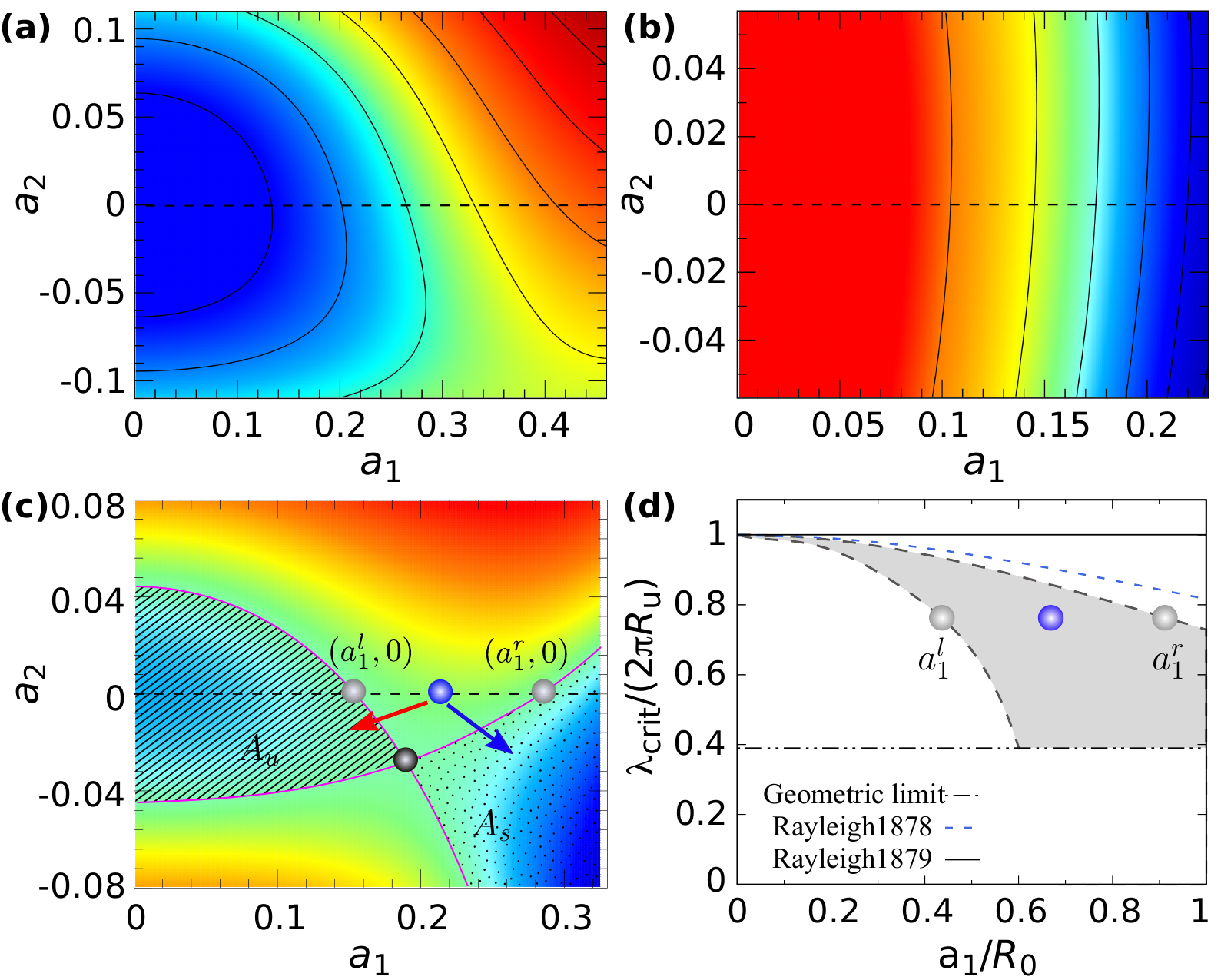}
	\caption{Surface area landscape. (a), (b), (c) 
	Contour plots of three typical surface area landscapes for small
	($\lambda=1$), large ($\lambda=4$)
	and intermediate ($\lambda=2$) wavelengths, respectively,  
	for a constant volume $V=1$.
	(d) The scaled critical wavelength for the breakup 
	versus the normalized amplitude
		from the present surface-area-minimization analysis. }
	\label{fig:2}
\end{figure}

Fig.~\ref{fig:2}(a), (b) and (c) illustrate
three typical surface areas $A$ as a function of the leading amplitude $a_1$ and the secondary amplitude $a_2$ for small, large, and intermediate wavelengths, respectively, for a constant volume $V=1$.
The effect of the higher order amplitudes $a_i$, $i\geq 3$ is minor~\cite{Note1}.
In each image, the red and blue regions represent 
higher and smaller values of the surface area, respectively.
For a small wavelength (Fig.~\ref{fig:2}(a)), 
the surface area has a global minimum 
at the position of $a_1=0$ and $a_2=0$ (dark blue region).
Hence, for any perturbations, the jet has to evolve to the dark blue 
region to minimize the surface area.
This kind of evolution is achieved by decreasing all the amplitudes till zero
and the final state of the jet is a uniform-radius cylinder.
Contrarily, for a large wavelength (Fig.~\ref{fig:2}(b)), 
the global minimum of the surface area (dark blue region) 
locates nearby the maximal value of the leading amplitude $a_1$.
Therefore, for any perturbations,
in order to reduce the surface area, the leading amplitude $a_1$ has
to continuously increase with time, which leads to the formation of droplets.

It is highlighted that for an intermediate wavelength, 
the surface area has two local minima, as portrayed in Fig.~\ref{fig:2}(c).
One minimum is at the position of $a_1=0$ and $a_2=0$, 
which is inside the region $A_u$ that is filled by hatch lines. 
The other one locates inside the region $A_s$ that is filled by small dots.
The boundaries of the regions $A_u$ and $A_s$ are marked
by the magenta lines, which are the 
isolines of the surface area of the saddle point (the black circle)
that is given by the locus of $\partial_{a_1} A=0$ 
and $\partial_{a_2} A=0$.
According to the surface-area-minimization principle,
the evolution behavior of the jet with a surface area landscape shown in Fig.~\ref{fig:2}(c)  is classified into three categories:
(i) For any perturbations inside the region $A_u$,
the jet cannot surmount the energy barrier at the boundary
(the curved magenta lines)  and has to evolve 
to the left local minimum in $A_u$, which leads to the formation of a uniform-radius jet.
(ii) Similarly, 
for any perturbations inside the region $A_s$,
the jet can also not overcome the energy barrier at the boundary 
and the leading amplitude $a_1$ has to 
increase with time to converge to the right local minimum in $A_s$,
which gives rise to a droplet-structure.
(iii) For perturbations outside the regions $A_u$ and $A_s$, 
the surface area of the jet is greater than the value at the saddle point
and the jet can evolve  either to the left local minimum in $A_u$ or 
to the right local minimum in $A_s$. 
The concrete evolution path depends upon the kinetics which is analyzed in the following.

In order to compare the present results with Rayleigh's criterion, 
we analyze the evolution behavior of the jet
with zero secondary amplitude as Rayleigh's consideration.
The horizontal dashed line ($a_2=0$) in Fig.~\ref{fig:2}(c) 
intersects with the isolines of the surface area of the saddle point 
at the points: $(a_1^l, 0)$ and $(a_1^r, 0)$, as represented by the gray circles.
According to the evolution behavior (iii),
a perturbed jet, $r=R_0+a_1\cos kz$ with $a_1^l \leq a_1 \leq a_1^r$, as 
exemplarily indicated 
by the blue circle in Fig.~\ref{fig:2}(c), 
can evolve either into a uniform-radius cylinder
by converging to the local minimum in $A_u$ (the red arrow) 
or into a sequence of droplets by converging to the local minimum in $A_s$ (the blue arrow). 
It is emphasized that such an interval 
$[a_1^l,a_1^r]$ is not only obtained for $\lambda=2$
that is illustrated in Fig.~\ref{fig:2}(c), but also
appears for many other intermediate wavelengths.

Fig.~\ref{fig:2}(d) illustrates all the intermediate wavelengths
which give rise to such an interval  $[a_1^l, a_1^r]$ 
as a function of the boundaries of the interval 
$a_1^l$ and $a_1^r$, as shown by the gray shaded region between the dashed lines. 
Here, the wavelength and the amplitude have been scaled by the perimeter $2\pi R_u$
and the radius $R_0$, respectively.
The two gray circles correspond to the ones which 
are shown in Fig.~\ref{fig:2}(c) for $\lambda=2$.
All the setups below the shaded region 
have a surface area landscape as in  Fig.~\ref{fig:2}(a)
and the final state is a uniform-radius cylinder.
All the setups above the shaded region 
have a surface area landscape as in  Fig.~\ref{fig:2}(b),
where the jet has to break into droplets. 
The critical setup for the breakup has to be inside 
the shaded region in order to obey the surface-area-minimization principle.
Rayleigh's criteria $\lambda_{\text{crit}}=2\pi R_0$~\cite{Rayleigh1878} 
and $\lambda_{\text{crit}}=2\pi R_u$~\cite{Rayleigh1879} 
are denoted by the blue dashed and the horizontal solid lines, respectively.
On the one hand, Rayleigh's criteria are based on the 
assumption of $a_1/R_0\ll 1$.
As can be seen in the region $a_1/R_0\ll 1$
in Fig.~\ref{fig:2}(d),  the present work is in good agreement 
with Rayleigh's criteria.
On the other hand,  Rayleigh's criteria deviate
from the gray shaded region when $a_1\sim R_0$.
The horizontal dot dashed line in Fig.~\ref{fig:2}(d) is obtained from 
the fact that the distance between the resulting droplets
cannot exceed the wavelength of the perturbation.
According to this geometric constraint, we obtain
a critical wavelength for the breakup
$\lambda_{\text{crit}}=\sqrt{6}R_u$.
After normalization, we have $\lambda_{\text{crit}}/2\pi R_u\approx0.39$,
as depicted by the horizontal dot dashed line in Fig.~\ref{fig:2}(d).

\begin{figure}[b]
	\centering
	\includegraphics[width=1\linewidth]{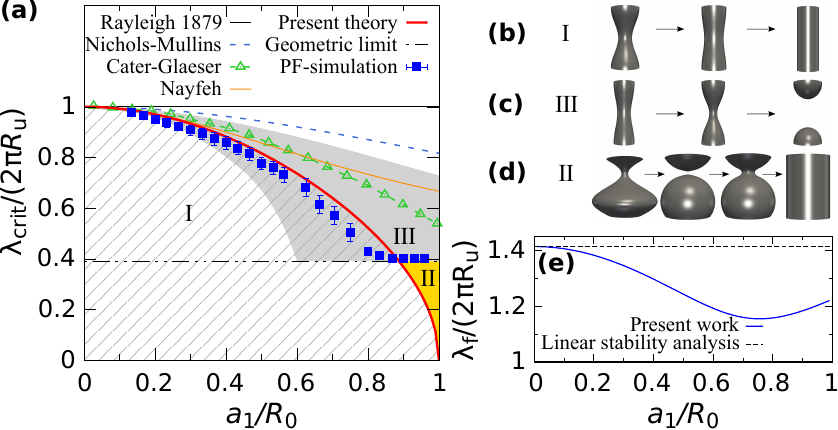}
	\caption{Stability diagram. (a) The scaled critical breakup wavelength   versus the normalized amplitude from the present work and literature.  (b), (c) and (d) Evolution behavior of liquid jets in the regions I, III and II
from the phase-field (PF) simulations.  
(e) Illustration of the wavelength, which refers to the fastest breakup rate, 
as a function of the amplitude, in comparison with the one from the linear stability analysis.
	}
	\label{fig:3}
\end{figure}

In contrast to the linear stability analysis~\cite{nichols1965} which 
gives rise to Rayleigh's criteria,
a high order stability analysis is carried out to 
derive the time evolution equation for the leading amplitude~\cite{Note1}, reading
\begin{align}
	\partial_ta_1&=-C\bigg\{k^2-\frac{1}{R_0^2}\bigg[1+\sum_{n=1}^\infty (na_nk)^2 \bigg]\bigg\}a_1k^2,
	\label{eq:19}
\end{align}
where $C$ is a prefactor.
When the leading amplitude 
increases with time ($\partial_ta_1>0$),
the jet transforms into droplets.
When the  leading amplitude 
decreases with time ($\partial_ta_1<0$),
the jet is morphologically stable.
Thus, the morphological stability criterion is given by the locus of $\partial_t a_1=0$,
which yields the most important finding of the present work
\begin{equation}
	\lambda_{\text{crit}}=2\pi \sqrt{R_0^2-\sum_{n=1}^\infty (na_n)^2}.
	\label{eq:criterion}
\end{equation}
It is noted that the stability criterion, Eq.~\eqref{eq:criterion} can 
be applied for a jet which is described by an infinite series 
as shown by Eq.~\eqref{eq:1}.
For a zero secondary amplitude as Rayleigh's consideration,
the stability criterion is expressed as 
$\lambda_{\text{crit}}=2\pi \sqrt{R_0^2-a_1^2}$.
An intuitive interpretation of this criterion 
is that when $\lambda>2\pi \sqrt{R_0^2-a_1^2}$,
the mean curvature at the trough of a wave-deformation
is higher than the one at the crest
and the induced flow/diffusion from the trough to the crest results in 
the breakup of the jet. The normalized critical wavelength 
$\lambda_{\text{crit}}/(2\pi R_u)$ versus the  scaled amplitude $a_1/R_0$
is  shown by  the red solid line in  Fig.~\ref{fig:3}(a).
The gray shaded region corresponds to the one as shown in Fig.~\ref{fig:2}(d).
As aforementioned, the critical wavelength has to be inside this shaded region 
to obey the surface-area-minimization principle.
Rayleigh's criteria 
$\lambda_{\text{crit}}=2\pi R_0$~\cite{Rayleigh1878} 
which is consistent with the linear stability analysis 
of Nichols-Mullins~\cite{nichols1965}
and $\lambda_{\text{crit}}=2\pi R_u$~\cite{Rayleigh1879} 
are depicted by the blue dashed  and horizontal solid lines, respectively.
The green dashed line with triangle symbols corresponds
to the results from Carter and Glaeser~\cite{carter1987}. 
This criterion is limited by the assumption that 
the surface of the jet can always be depicted by 
a single cosine function during all the time.
The orange solid line shows the criterion which is 
derived by Nayfeh~\cite{nayfeh1970nonlinear} using a second order stability analysis. 

There are two highlighted features of the 
present stability criterion. Firstly, the present stability criterion is inside the 
shaded region and does obey the physical principle of free energy minimization
for all the possible amplitudes $0\leq a_1\leq R_0$.
Secondly and most importantly, 
in contrast to other criteria, the present stability criterion crosses with the 
horizontal dot dashed line (the geometric limit)
and the stability diagram is divided into three regions I (hatch lines), 
II (orange area) and III (other areas).
For all the setups in the regions I and III, 
the end-state is a uniform-radius jet and a sequence of droplets, 
as shown in Fig.~\ref{fig:3}(b) and (c),  respectively.
The evolution behavior in the region II is determined by two facts:
one fact is the stability criterion given by
the red solid line
according to which
the jet must transform into droplets. The other fact
is that the region II is below the line of the geometric constraint, 
according to which the final state of the jet  is a uniform-radius cylinder.
These two facts are actually not in conflict with each other, 
but rather can take place in a sequential manner, as confirmed by the phase-field 
simulation~\cite{Note1}  shown in  Fig.~\ref{fig:3}(d).
For perturbed jets in the region II, 
the jet firstly breaks up into several droplets because the setup is above the red solid line in Fig.~\ref{fig:3}(a).
But, the diameter of the resulting droplets
is greater than the wavelength
and hence, the neighboring droplets get in touch with each other, resulting in reforming a 
uniform-radius jet.
This natural phenomenon that the reformation of a uniform-radius
jet after the breakup is due to the fact that the present stability criterion 
intersects with the geometric constraint in Fig.~\ref{fig:3}(a).
We emphasize that all other criteria in literature do not cross with the 
geometric constraint and therefore cannot address this natural phenomenon.  
The intersection of the stability criterion with the geometric limit is also
clearly confirmed by the phase-field simulations~\cite{wang2012}, 
as shown by the square symbols in Fig.~\ref{fig:3}(a).

\begin{figure}[b]
	\centering
	\includegraphics[width=0.99\linewidth]{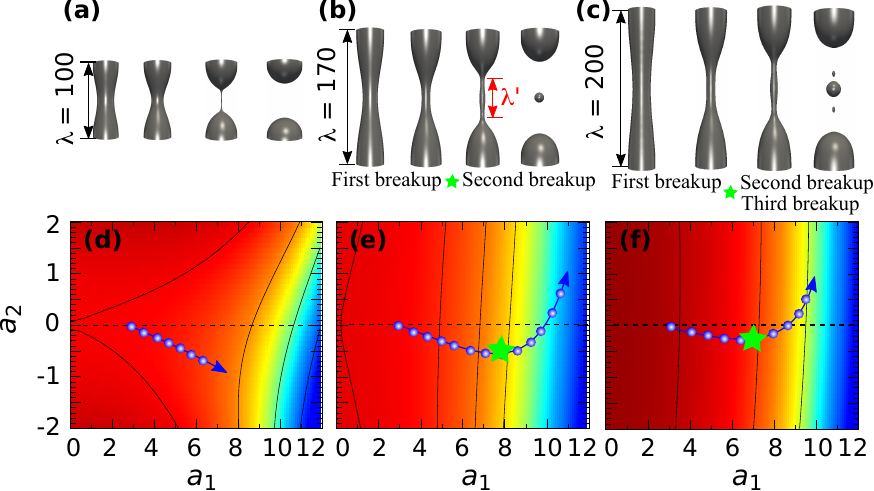}
	\caption{Satellite droplets.  (a), (b) and (c) Phase-field 
	simulations of the morphological evolution of jets in the morphological
	unstable region with different wavelengths.  
	(d), (e) and (f) Evolution path of the liquid jets in (a), (b) and (c), respectively.}
	\label{fig:4}
\end{figure}

The experimentally observed breakup may occur 
at the wavelength  $\lambda_{\text{f}}$ which refers
to the fastest breakup rate, rather than at the
critical breakup wavelength.
The fastest breakup wavelength is  straightforwardly
 obtained by the condition
of $\partial_k(\partial_ta_1)=0$ based on Eq.~\eqref{eq:19}.
The normalized fastest breakup wavelength  $\lambda_{\text{f}}/(2\pi R_u)$ 
versus the scaled amplitude $a_1/R_0$
is illustrated in Fig.~\ref{fig:3}(e) by the blue solid line.
The horizontal dashed line depicts the
fastest breakup wavelength  $\lambda_{\text{f}}=\sqrt{2} (2\pi R_u)$
from the linear stability analysis \cite{nichols1965,Rayleigh1879}.
The present work coincides with the linear stability analysis in the region of $a_1/R_0\ll 1$
and with an increase of the amplitude, the deviation becomes larger.

Fig.~\ref{fig:4}(a)-(c) illustrate the time evolution of three jets 
initially with $R_0=15$, $a_1=3$, $a_i=0$, $i\geq2$,
but different wavelengths from the phase-field simulations.  
These three setups are in the morphologically unstable region in Fig.~\ref{fig:3}(a) 
and the final states are droplet-structures.
For $\lambda=100$,  the jet breaks up into a sequence of uniform-radius droplets.
For $\lambda=170$ and $\lambda=200$,
mini-droplets appear  between the main droplets.
In order to understand these three different
breakup behaviors in Fig.~\ref{fig:4}(a)-(c),
we trace the evolution path of the jet by 
a Fourier decomposition to analyze the surface of jets
to obtain the leading amplitude 
$a_1$ and the secondary amplitude $a_2$.
The obtained values
$(a_1, a_2)$ at different time steps from the phase-field simulations, 
as shown by the blue circles in Fig.~\ref{fig:4}(d)-(f),
are illustrated in the contour plot of the surface area landscape. 
The blue arrow depicts the temporal evolution direction.
As can be seen from the evolution path in Fig.~\ref{fig:4}(d)-(f),
 the leading amplitude $a_1$  increases with time in all cases 
 to converge to a lower surface area state (dark blue region),
 which leads to the formation of a droplet-structure. 
The significant difference is the evolution behavior of the secondary 
amplitude. The secondary amplitude $a_2$ 
continuously decreases with time for $\lambda=100$,
whereas for $\lambda=170$ and $200$, the secondary amplitude  
first decreases and then increases with time after passing through the star point.

In the first case ($\lambda=100$),
the global minimum of the surface area
locates at the right bottom of the surface area landscape
and the isolines of the surface area (black solid lines)
are axi-asymmetric with respect to the line of $a_2=0$.
In order to reduce the surface area, the secondary amplitude 
continuously decreases with time, giving rise
to the formation of uniform-radius droplets. 
In the second and third cases ($\lambda=170, 200$),
the isolines of the surface area are almost axi-symmetric with respect to 
the line of  $a_2=0$ and the lower surface area (dark blue region) occurs
both at  positive and negative values of the secondary amplitude.
Before reaching the star point, 
the secondary amplitude decreases with time as the first case of $\lambda=100$.
At the star point,  a secondary breakup
with wavelength  $\lambda^\prime$ satisfying the breakup criterion takes place,
which results in an increase of the secondary amplitude.
It is noted that in all the cases in Fig.~\ref{fig:4} irrespective of
of the formation of satellites,
the evolution of the jet follows the principle of surface area minimization.
When the surface area landscape is asymmetric
with respect to the line of $a_2=0$,
the breakup is regular and a chain of uniform-radius droplets is formed.
When the surface area landscape is symmetric
with respect to the line of $a_2=0$,
which allows to
decrease the surface area by increasing the secondary amplitude,
a second breakup happens and satellite droplets appear.
Hence,  it is capable to control the formation 
of satellite droplets by  manipulating
the surface area landscape.

In conclusion, we have shed new light on the formation of droplets
and mini-droplets  from liquid jets by proposing a novel approach analyzing
the surface area landscape.
 We have also addressed a new phenomenon that the jet initially transforms into droplets and then 
 becomes a  uniform-radius cylinder again. The current results can 
 be applied for topics where liquid jets are present.

%  
% \bibliographystyle{apsrev4-1}
% \bibliography{reference_Wang}
\end{document}